\documentclass{aastex62}

\received{}
\revised{}
\accepted{August 27, 2019}
\submitjournal{ApJ}

\shorttitle{$^{13}$C Isotopic Fractionation of CCH in Two Starless Cores: L1521B and L134N}
\shortauthors{Taniguchi et al.}

\begin{document}

\title{Investigation of $^{13}$C Isotopic Fractionation of CCH in Two Starless Cores: L1521B and L134N}

\correspondingauthor{Kotomi Taniguchi}
\email{kotomi.taniguchi@gakushuin.ac.jp}

\author[0000-0003-4402-6475]{Kotomi Taniguchi}
\altaffiliation{Virginia Initiative on Cosmic Origins Fellow}
\altaffiliation{Current institution: Department of Physics, Faculty of Science, Gakushuin University, Mejiro, Toshima, Tokyo 171-8588, Japan}
\affiliation{Department of Astronomy, University of Virginia, Charlottesville, VA 22904, USA}
\affiliation{Department of Chemistry, University of Virginia, Charlottesville, VA 22903, USA}

\author[0000-0002-4649-2536]{Eric Herbst}
\affiliation{Department of Astronomy, University of Virginia, Charlottesville, VA 22904, USA}
\affiliation{Department of Chemistry, University of Virginia, Charlottesville, VA 22903, USA}

\author[0000-0001-8231-9493]{Hiroyuki Ozeki}
\affiliation{Department of Environmental Science, Faculty of Science, Toho University, Miyama, Funabashi, Chiba 274-8510, Japan}

\author[0000-0003-0769-8627]{Masao Saito}
\affiliation{National Astronomical Observatory of Japan (NAOJ), Osawa, Mitaka, Tokyo 181-8588, Japan}
\affiliation{Department of Astronomical Science, School of Physical Science, SOKENDAI (The Graduate University for Advanced Studies), Osawa, Mitaka, Tokyo 181-8588, Japan}



\begin{abstract}

We have carried out observations of CCH and its two $^{13}$C isotopologues, $^{13}$CCH and C$^{13}$CH, in the 84 -- 88 GHz band toward two starless cores, L1521B and L134N (L183), using the Nobeyama 45 m radio telescope.
We have detected C$^{13}$CH with a signal-to-noise (S/N) ratio of 4, whereas no line of $^{13}$CCH was detected in either the dark clouds.
The column densities of the normal species were derived to be ($1.66 \pm 0.18$)$\times 10^{14}$ cm$^{-2}$ and ($7.3 \pm 0.9$)$\times 10^{13}$ cm$^{-2}$ ($1 \sigma$) in L1521B and L134N, respectively.
The column density ratios of $N$(C$^{13}$CH)/$N$($^{13}$CCH) were calculated to be $>1.1$ and $>1.4$ in L1521B and L134N, respectively.
The characteristic that $^{13}$CCH is less abundant than C$^{13}$CH is likely common for dark clouds.
Moreover, we find that the $^{12}$C/$^{13}$C ratios of CCH are much higher than those of HC$_{3}$N in L1521B by more than a factor of 2, as well as in Taurus Molecular Cloud-1 (TMC-1).
In L134N, the differences in the $^{12}$C/$^{13}$C ratios between CCH and HC$_{3}$N seem to be smaller than those in L1521B and TMC-1.
We discuss the origins of the $^{13}$C isotopic fractionation of CCH and investigate possible routes that cause the significantly high $^{12}$C/$^{13}$C ratio of CCH especially in young dark clouds, with the help of chemical simulations.
The high $^{12}$C/$^{13}$C ratios of CCH seem to be caused by reactions between hydrocarbons (e.g., CCH, C$_{2}$H$_{2}$, $l,c$-C$_{3}$H) and C$^{+}$.
\end{abstract}



\section{Introduction} \label{sec:intro}

Exotic unsaturated carbon-chain molecules are one of the crucial constituents of approximately 200 molecules detected in the interstellar medium (ISM) and circumstellar shells.
In fact, they account for around 40\% of the interstellar molecules.
Therefore, it is important for astrochemists to understand carbon-chain chemistry.

These carbon-chain species have long been associated with young starless cores such as Taurus Molecular Cloud-1 \citep[TMC-1;][]{1992ApJ...392..551S,2004PASJ...56...69K}.
They are formed by gas-phase ion-molecule reactions and neutral-neutral reactions before carbon atoms are converted into CO molecules.
Besides the classical carbon-chain chemistry, an ion-molecule chemistry occurring at somewhat higher temperatures, and starting from gaseous methane (CH$_{4}$), named warm carbon-chain chemistry \citep[WCCC;][]{2013ChRv..113.8981S}, was found to occur  around low-mass Class 0/I protostars such as L1527.  
In particular, methane sublimated from dust grains reacts with ionic carbon (C$^{+}$) in the gas phase, which is a trigger of successive carbon-chain formation.
It was recently found that formation of cyanopolyynes (HC$_{2n+1}$N, $n=1,2,3,...$) occurs in the warm dense gas around high-mass protostellar objects \citep{2018ApJ...854..133T,2019ApJ...872..154T}. 

The formation and destruction mechanisms of carbon-chain molecules were investigated mainly by astrochemical simulations \citep[e.g.,][]{1992ApJ...392..551S}.
In these early stages, it was unclear what specific reactions significantly contribute to the formation of carbon-chain species.
Another method to investigate the main formation pathways of carbon-chain molecules consists of observations of the $^{13}$C isotopic fractionation \citep[e.g.,][]{1998AA...329.1156T}.

The first hint of  $^{13}$C isotopic fractionation for carbon-chain molecules was found in HC$_{5}$N toward TMC-1 using the Nobeyama 45 m radio telescope \citep{1990ApJ...361L..15T}.
However, signal-to-noise ratios were not high enough to confirm the differences in abundances among its five $^{13}$C isotopologues and such studies were left for future work.
The confirmation of the $^{13}$C isotopic fractionation was achieved for HC$_{3}$N in TMC-1 \citep{1998AA...329.1156T}.
Other observations including fractionation studies targeting different carbon-chain molecules were carried out in TMC-1. 
These species include CCS \citep{2007ApJ...663.1174S}, CCH \citep{2010AA...512A..31S}, C$_{3}$S and C$_{4}$H \citep{2013JPCA..117.9831S}, HC$_{5}$N \citep{2016ApJ...817..147T}, and HC$_{7}$N \citep{2018MNRAS.474.5068B}.

Based on the observations mentioned in the preceding paragraph, some possible main formation pathways of cyanopolyynes were investigated \citep{1998AA...329.1156T,2016ApJ...817..147T,2018MNRAS.474.5068B}.
In the case of HC$_{3}$N, the abundances of H$^{13}$CCCN and HC$^{13}$CCN are similar to each other, and HCC$^{13}$CN is more abundant than the others.
From the results, the reaction between C$_{2}$H$_{2}$ and CN was suggested as the main formation pathway of HC$_{3}$N \citep{1998AA...329.1156T}.
On the other hand, there is no significant difference in abundance among the five $^{13}$C isotopologues of HC$_{5}$N.
Reactions between hydrocarbon ions (C$_{5}$H$_{n}^{+}$, $n=3,4,5$) and nitrogen atoms followed by  dissociative recombination reactions were found to be the most plausible route to explain the observed $^{13}$C isotopic fractionation of HC$_{5}$N \citep{2016ApJ...817..147T}.
These proposed reactions were supported by the $^{14}$N/$^{15}$N ratios of HC$_{3}$N and HC$_{5}$N \citep{2017PASJ...69L...7T}.
In the case of HC$_{7}$N, the fractionation results and proposed main formation mechanism are similar to those of HC$_{5}$N \citep{2018MNRAS.474.5068B}.

The main formation mechanisms of HC$_{3}$N were investigated in other sources including  the L1527 low-mass star-forming core and the G28.28--0.36 high-mass star-forming core \citep{2016ApJ...830..106T}, as well as two starless cores \citep[L1521B and L134N;][]{2017ApJ...846...46T}.
Except for the case of L134N, the reaction between C$_{2}$H$_{2}$ and CN has been proposed as its main formation mechanism \citep{2016ApJ...830..106T,2017ApJ...846...46T}, while  the reaction between CCH and HNC could explain the observed $^{13}$C isotopic fractionation in L134N \citep{2017ApJ...846...46T}.
The proposed main formation pathway of HC$_{3}$N in the star-forming cores L1527 and G28.28--0.36 is consistent with model calculations for WCCC \citep{2008ApJ...681.1385H} and for hot cores \citep{2019arXiv190611296T}.
The differences among starless cores are probably caused by their different ages; L134N is considered to be more evolved than L1521B and TMC-1 \citep{2017ApJ...846...46T}.

Another interesting feature involving carbon isotopes was the observation that the $^{12}$C/$^{13}$C ratios of carbon-chain molecules are higher than the elemental ratio in the local interstellar medium \citep[$60-70$;][]{2005ApJ...634.1126M}, an effect known as the dilution of the $^{13}$C species\footnote{In this paper, we define the dilution of the $^{13}$C species as the $^{12}$C/$^{13}$C ratios higher than 70, which is the mean $^{12}$C/$^{13}$C ratio in the local interstellar medium.}.
This dilution is considered to be caused at least in part by the low $^{13}$C$^{+}$ abundance which occurs via the  reaction:
\begin{equation} \label{rea:co}
^{13}{\rm {C}}^{+} + {\rm {CO}} \rightarrow {\rm {C}}^{+} + ^{13}\!{\rm {CO}} + \Delta E \; (35 \; {\rm {K}}),
\end{equation}
a reaction that is efficient especially in low-temperature conditions \citep{1984ApJ...277..581L}.
The importance of this reaction stems from the fact that the initial step in the formation of carbon-chain molecules in dark clouds occurs via gas-phase ion-molecule reactions or neutral-neutral reactions with C$^{+}$ or C.
Hence, the loss of the $^{13}$C$^{+}$ abundance leads to the high $^{12}$C/$^{13}$C ratios of carbon-chain molecules.
However, the different degrees of the dilution of the $^{13}$C species among carbon-chain molecules found in TMC-1 cannot  be explained only by reaction (\ref{rea:co}) \citep{2016ApJ...817..147T}.

In this paper, we report the observations of the $N=1-0$ transition lines of CCH and its two $^{13}$C isotopologues in L1521B ($d=140$ pc)  and L134N ($d=110$ pc) using the Nobeyama 45 m radio telescope.
We describe our observations in Section \ref{sec:obs}.
The results and derived parameters with the methods utilized are presented in Section \ref{sec:resana}.
The differential fractionation between the two $^{13}$C-containing CCH isotopologues in L1521B and L134N and possible mechanisms causing the heavy dilution of $^{13}$C-containing species especially of CCH in dark clouds are discussed with the help of a chemical simulation in Sections \ref{sec:dis1} and \ref{sec:dis2}, respectively.
Our conclusions are summarized in Section \ref{sec:con}.

\section{Observations} \label{sec:obs}

The observations were carried out in 2019 January with the Nobeyama 45-m radio telescope (Proposal ID: CG181003, PI: Kotomi Taniguchi, 2018-2019 season).
The $N=1-0$ transition lines of CCH and its two $^{13}$C isotopologues in the 84 -- 88 GHz band were observed simultaneously with the T70 receiver.
The beam size and main beam efficiency ($\eta_{\rm {mb}}$) were 19\arcsec and 55\%, respectively.
The system temperatures were between 170 and 270 K depending on the weather conditions and elevation.
We used the SAM45 FX-type digital correlator in the frequency setup whose bandwidth and frequency resolution were 125 MHz and 30.52 kHz, respectively.
The frequency resolution corresponds to the velocity resolution of 0.1 km s$^{-1}$ at 86 GHz.
We conducted the 2-channel binning in the final spectra, and thus the velocity resolution of the final spectra is 0.2 km s$^{-1}$.

The position-switching mode was employed.
The observed positions were ($\alpha_{2000}$, $\delta_{2000}$) = (04$^{\rm h}$24$^{\rm m}$12\fs67, +26\arcdeg36\arcmin52\farcs8) and (15$^{\rm h}$54$^{\rm m}$12\fs72, -02\arcdeg49\arcmin47\farcs4) for L1521B and L134N, respectively.
The off position for L1521B was set to be ($\Delta \alpha$, $\Delta \delta$) = (+4\arcmin, +4\arcmin) away from the on-source position, and that for L134N was set at $+3\arcmin$ away in the right ascension.
The scan pattern was 20 s and 20 s for on-source and off-source positions, respectively.
The chopper-wheel calibration method was adopted and hence the absolute calibration error was approximately 10\%.

We checked the pointing accuracy by observations of the SiO ($J=1-0$) maser lines from NML Tau at ($\alpha_{2000}$, $\delta_{2000}$) = (03$^{\rm h}$53$^{\rm m}$28\fs86, +11\arcdeg24\arcmin22\farcs4) and WX-Ser at ($\alpha_{2000}$, $\delta_{2000}$) = (15$^{\rm h}$27$^{\rm m}$47\fs05, +19\arcdeg33\arcmin51\farcs8) during the observations of L1521B and L134N, respectively.
The pointing observations were conducted using the H40 receiver every 1.5 hour.
The pointing accuracy was within 3\arcsec.

\section{Results and Analyses} \label{sec:resana}

\subsection{Results}

\floattable
\begin{deluxetable}{lcccccccc}
\tabletypesize{\scriptsize}
\tablecaption{Spectral line parameters of the $N=1-0$ transition of CCH and its two $^{13}$C isotopologues in L1521B and L134N \label{tab:t1}}
\tablewidth{0pt}
\tablehead{
\colhead{Species} & \colhead{Transition} & \colhead{Frequency\tablenotemark{a}}  & \colhead{$S$\tablenotemark{a,b}} & \colhead{$T_{\rm {mb}}$\tablenotemark{c}} & \colhead{$\Delta v$\tablenotemark{c}} & \colhead{$V_{\rm {LSR}}$\tablenotemark{d}} & \colhead{$\int T_{\rm {mb}}dv$\tablenotemark{c}} & \colhead{rms\tablenotemark{e}} \\
\colhead{ }  &  \colhead{ }   & \colhead{(GHz)}  & \colhead{ }  &  \colhead{(K)} & \colhead{(km s$^{-1}$)} & \colhead{(km s$^{-1}$)} & \colhead{(K km s$^{-1}$)} & \colhead{(mK)}
}
\startdata
 {\bf {L1521B}} & & & & & & & & \\
     CCH & $J=3/2-1/2, F= 1- 1$ & 87.284156 & 0.17 & 0.588 (9) & 0.423 (8) & 6.5 & 0.265 (6) & 4.7 \\ 
             & $J=3/2-1/2, F= 2- 1$ & 87.316925 & 1.67 & 1.165 (12) & 0.459 (5) & 6.4 & 0.569 (9) & 4.7 \\
             & $J=3/2-1/2, F= 1- 0$ & 87.328624 & 0.83 & 0.843 (10) & 0.4223 (6) & 6.5 & 0.379 (7) & 4.7 \\
             & $J=1/2-1/2, F= 1- 1$ & 87.402004 & 0.83 &1.13 (2) & 0.405 (9) & 6.4 & 0.485 (14) & 5.6 \\
             & $J=1/2-1/2, F= 0- 1$ & 87.407165 & 0.33 & 0.865 (15) & 0.408 (8) & 6.4 & 0.375 (10) & 5.6 \\
             & $J=1/2-1/2, F= 1- 0$ & 87.446470 & 0.17 & 0.678 (12) & 0.405 (8) & 6.4 & 0.292 (8) & 5.6 \\ 
     $^{13}$CCH\tablenotemark{f} & $J=3/2-1/2, F_{1}= 2- 1, F=5/2-3/2$ & 84.119329 & 2.00 & $<0.016$ & ... & ... & $<0.007$ & 3.2 \\ 
             & $J=3/2-1/2, F_{1}= 2- 1, F=3/2-1/2$ & 84.124143 & 1.22 & $<0.016$ & ... & ... & $<0.007$ & 3.0 \\ 
             & $J=3/2-1/2, F_{1}= 1- 0, F=1/2-1/2$ & 84.151352 & 0.66 & $<0.016$ & ... & ... & $<0.007$ & 2.9\\             
     C$^{13}$CH\tablenotemark{f} & $J=3/2-1/2, F_{1}= 2- 1, F=5/2-3/2$ & 85.229326 & 2.00 & 0.026 (13) & 0.31 (19) & 6.5 & 0.0084 (7) & 3.8 \\
             & $J=3/2-1/2, F_{1}= 2- 1, F=3/2-1/2$ & 85.232792 & 1.25 & 0.025 (7) &  0.53 (16) & 6.5 & 0.0139 (6) & 3.8 \\
             & $J=3/2-1/2, F_{1}= 1- 0, F=1/2-1/2$ & 85.247708 & 0.65 & 0.016 (5) & 0.7 (3) & 6.4 & 0.0129 (6) & 3.5 \\ 
             & $J=3/2-1/2, F_{1}= 1- 0, F=3/2-1/2$ & 85.256952 & 1.28 & $<0.02$ & ... & ... & $<0.009$ & 3.6 \\       
      {\bf {L134N (L183)}} & & & & & & & & \\     
      CCH & $J=3/2-1/2, F= 1- 1$ & 87.284156 & 0.17 & 0.358 (11) & 0.311 (12) & 2.6 & 0.118 (6) & 4.8 \\
              & $J=3/2-1/2, F= 2- 1$ & 87.316925 & 1.67 & 0.853 (8) & 0.379 (4) & 2.5 & 0.344 (5) & 4.8 \\
              & $J=3/2-1/2, F= 1- 0$ & 87.328624 & 0.83 & 0.637 (12) & 0.339 (8) & 2.5 & 0.229 (7) & 4.8 \\
              & $J=1/2-1/2, F= 1- 1$ & 87.402004 & 0.83 & 0.777 (11) & 0.335 (5) & 2.5 & 0.277 (6) & 5.4 \\
              & $J=1/2-1/2, F= 0- 1$ & 87.407165 & 0.33 & 0.538 (16) & 0.320 (12) & 2.5 & 0.183 (9) & 5.4 \\
              & $J=1/2-1/2, F= 1- 0$ & 87.446470 & 0.17 & 0.415 (15) & 0.303 (14) & 2.5 & 0.134 (8) & 5.4 \\
     $^{13}$CCH\tablenotemark{f} & $J=3/2-1/2, F_{1}= 2- 1, F=5/2-3/2$ & 84.119329 & 2.00 & $<0.016$ & ... & ... & $<0.006$ & 3.1 \\
              & $J=3/2-1/2, F_{1}= 2- 1, F=3/2-1/2$ & 84.124143 & 1.22 & $<0.016$ & ... & ... & $<0.006$ & 3.1 \\
              & $J=3/2-1/2, F_{1}= 1- 0, F=1/2-1/2$ & 84.151352 & 0.66 & $<0.016$ & ... & ... & $<0.006$ & 2.9 \\
     C$^{13}$CH\tablenotemark{f} & $J=3/2-1/2, F_{1}= 2- 1, F=5/2-3/2$ & 85.229326 & 2.00 & 0.016 (6) & 0.7 (3) & 2.6 & 0.011 (6) & 3.4 \\
              & $J=3/2-1/2, F_{1}= 2- 1, F=3/2-1/2$ & 85.232792 & 1.25 & 0.020 (6) & 0.42 (14) & 2.5 & 0.009 (4) & 3.3 \\
              & $J=3/2-1/2, F_{1}= 1- 0, F=1/2-1/2$ & 85.247708 & 0.65 & $<0.02$ & ... & ... & $<0.007$ & 3.6 \\
              & $J=3/2-1/2, F_{1}= 1- 0, F=3/2-1/2$ & 85.256952 & 1.28 & $<0.02$ & ... & ... & $<0.007$ & 3.7 \\
\enddata
\tablenotetext{a}{Taken from the Cologne Database for Molecular Spectroscopy, CDMS \citep{2005JMoSt.742..215M}.}
\tablenotetext{b}{Intrinsic line strength.}
\tablenotetext{c}{The numbers in parentheses represent the standard deviation in the Gaussian fit. The errors are written in units of the last significant digit.}
\tablenotetext{d}{The errors were 0.2 km s$^{-1}$, corresponding to the velocity resolution of the final spectra.}
\tablenotetext{e}{The rms noises were evaluated in emission-free region in the $T_{\rm A}^{\ast}$ scale.}
\tablenotetext{f}{The upper limits of the peak intensities correspond to the $3\sigma$ limits and those of the integrated intensities were derived from the upper limits of peak intensities assuming that the line widths are equal to the average values of the normal species (0.42 km s$^{-1}$ and 0.33 km s$^{-1}$ in L1521B and L134N, respectively).}
\end{deluxetable}	

\begin{figure}[!h]
\figurenum{1}
 \begin{center}
  \includegraphics[width=12cm, bb= 0 0 388 692]{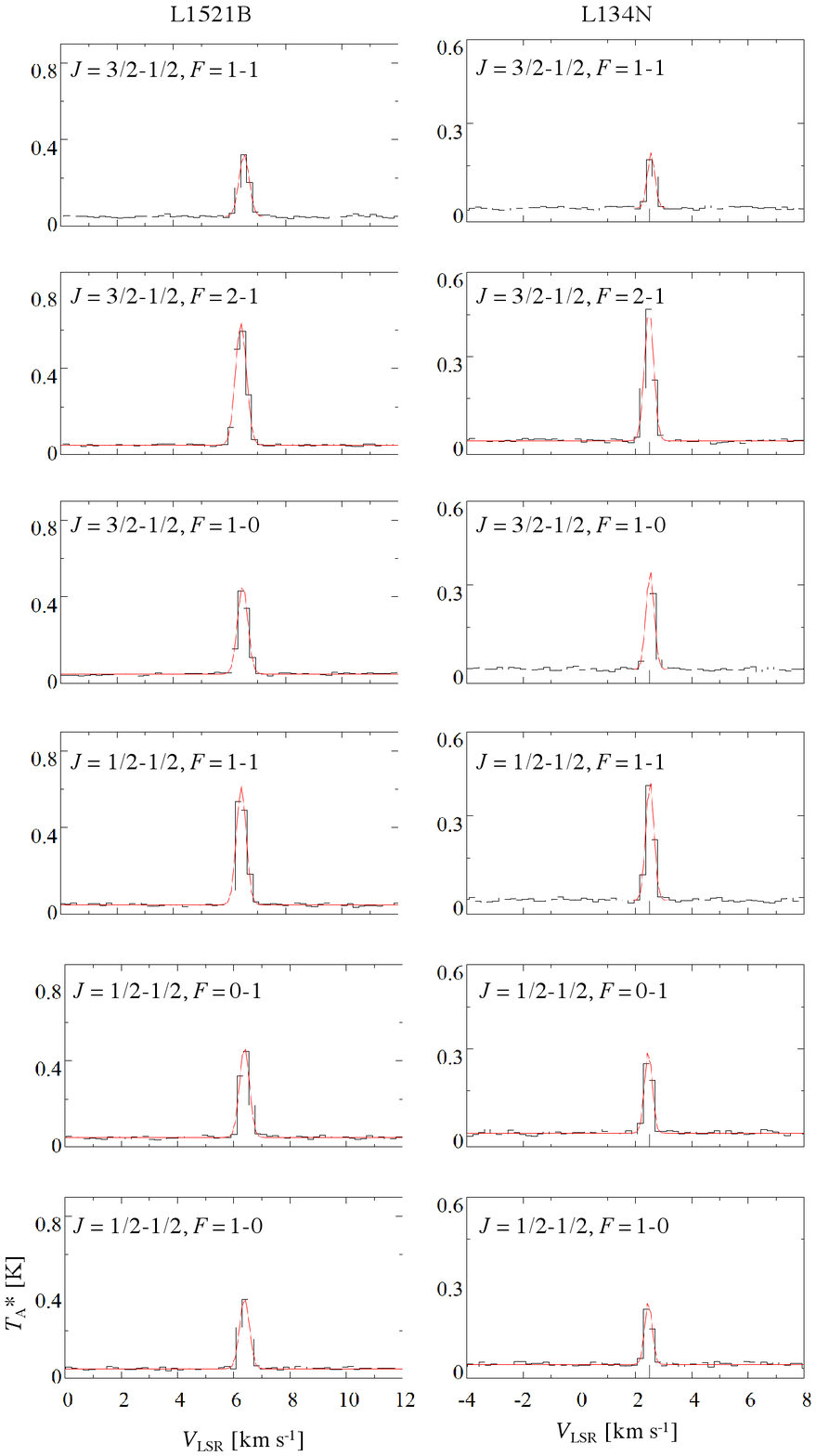}
 \end{center}
\caption{Spectra of the $N=1-0$ transition lines of CCH in L1521B and L134N. The vertical lines indicate the systemic velocities of each source (6.5 km s$^{-1}$ and 2.5 km s$^{-1}$ for L1521B and L134N, respectively). The red curves show the results of the best Gaussian fit.} \label{fig:f1}
\end{figure}

\begin{figure}[!h]
\figurenum{2}
 \begin{center}
  \includegraphics[width=11cm, bb= 0 0 386 462]{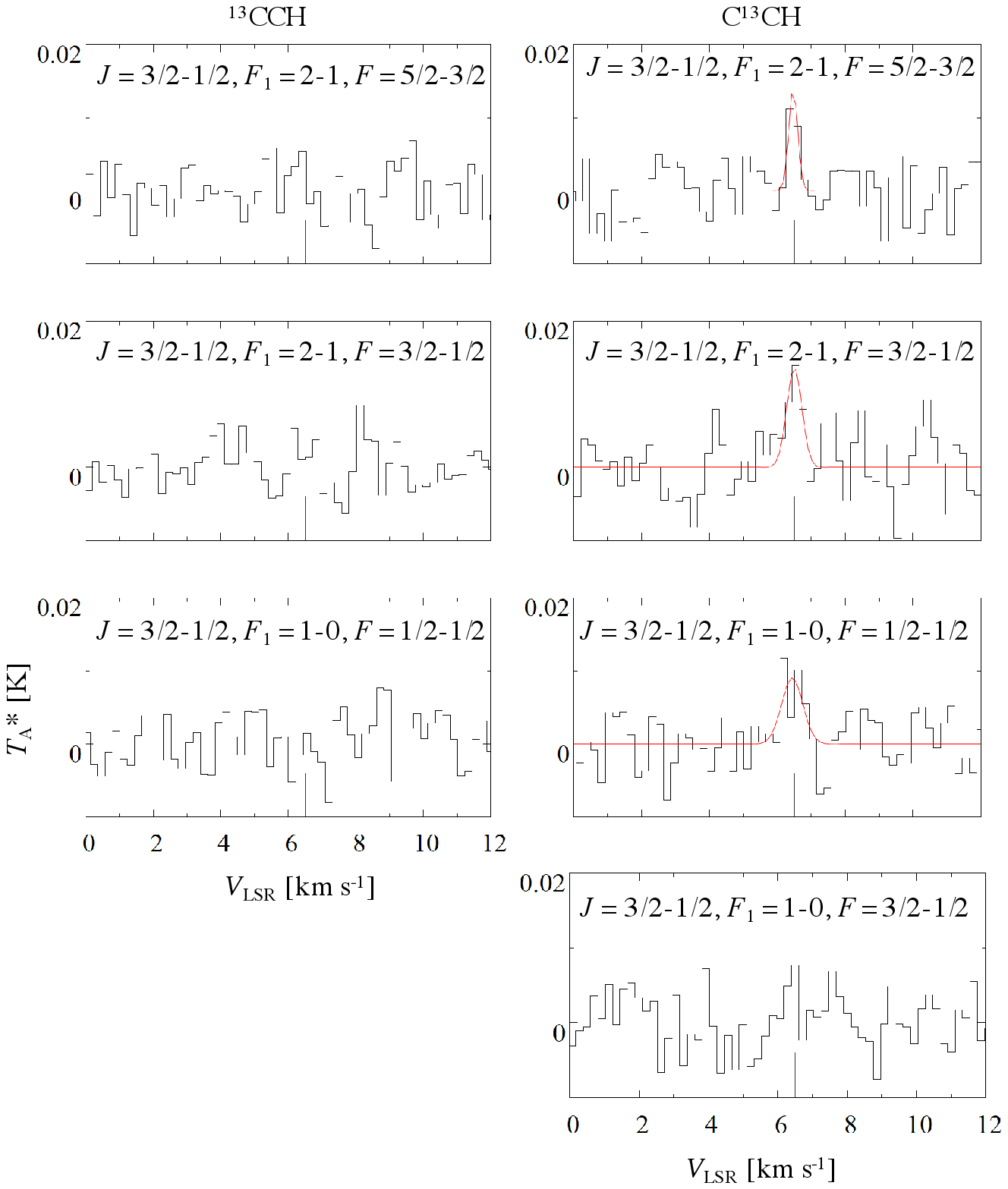} 
 \end{center}
\caption{Spectra of the $N=1-0$ transition lines of $^{13}$CCH and C$^{13}$CH in L1521B. The vertical lines indicate the systemic velocity (6.5 km s$^{-1}$). The red curves show the results of the best Gaussian fit.} \label{fig:f2}
\end{figure}

\begin{figure}[!h]
\figurenum{3}
 \begin{center}
  \includegraphics[width=11cm, bb= 0 0 392 465]{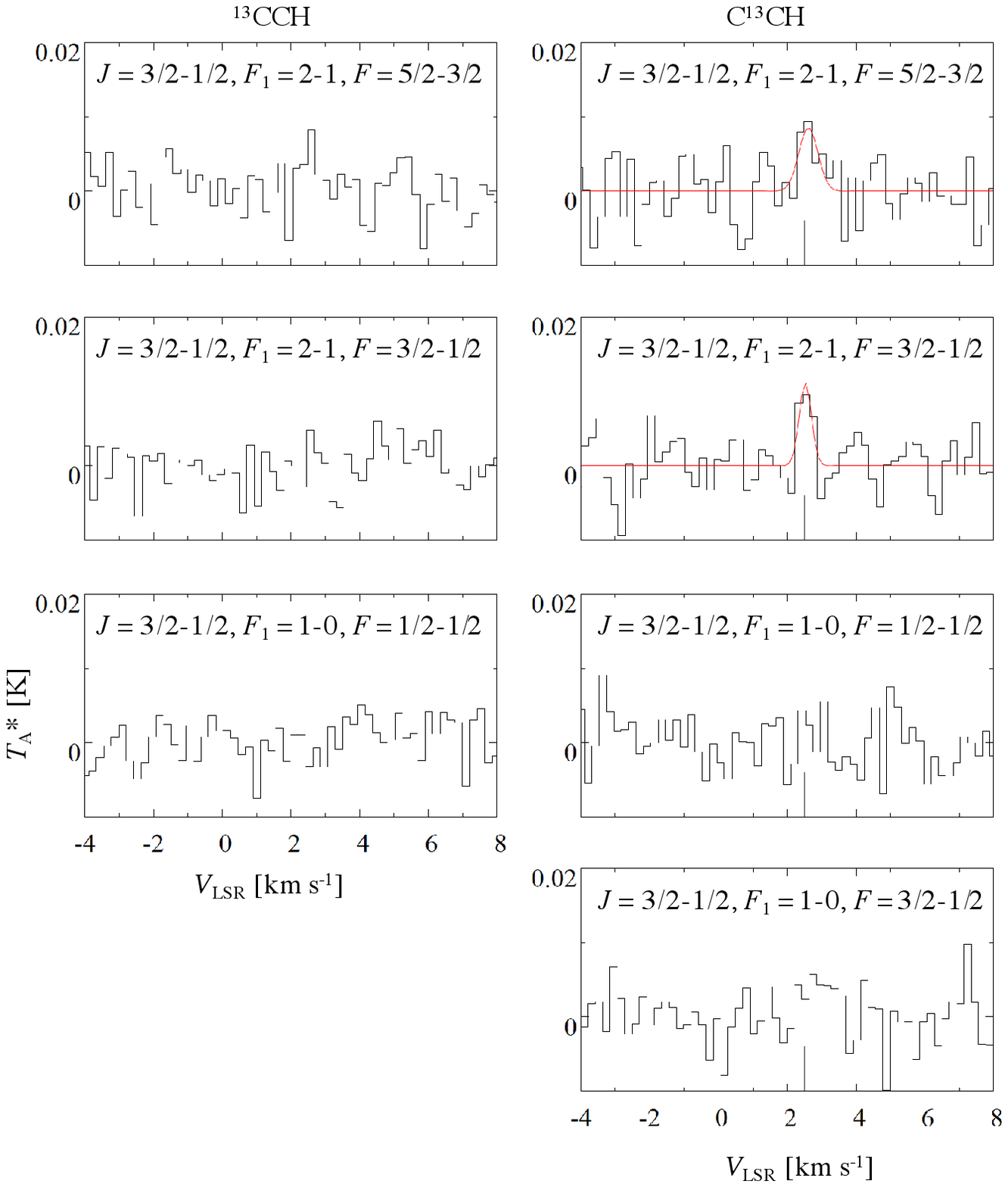} 
 \end{center}
\caption{Spectra of the $N=1-0$ transition lines of $^{13}$CCH and C$^{13}$CH in L134N. The vertical lines indicate the systemic velocity (2.5 km s$^{-1}$). The red curves show the results of the best Gaussian fit.} \label{fig:f3}
\end{figure}

We conducted the data reduction using Java NEWSTAR, which is the software for data reduction and analyses of the Nobeyama data.
The total on-source integration times were 18.75 hr and 23.5 hr for L1521B and L134N, respectively.
We fitted the spectra with a Gaussian profile,  and  the  obtained spectral line parameters are summarized in Table \ref{tab:t1}.

Figure \ref{fig:f1} shows the spectra of the normal species of CCH in L1521B and L134N.
The black vertical lines indicate the systemic velocities of each source, which are 6.5 km s$^{-1}$ and 2.5 km s$^{-1}$ in L1521B and L134N, respectively.
The velocity components of all the transition lines are consistent with the systemic velocities of each source within their errors of 0.2 km s$^{-1}$.  
Figures \ref{fig:f2} and \ref{fig:f3} show the spectra of the isotopomers $^{13}$CCH (left panels) and C$^{13}$CH (right panels) in L1521B and L134N, respectively.
In either the sources, no line of $^{13}$CCH was detected.
Two strong transition lines ($J=3/2-1/2, F_{1}= 2- 1, F=5/2-3/2$ and $J=3/2-1/2, F_{1}= 2- 1, F=3/2-1/2$) of C$^{13}$CH were detected in L1521B and L134N with a signal-to-noise (S/N) ratio of 4.
In addition, the weaker transition line ($J=3/2-1/2, F_{1}= 1- 0, F=1/2-1/2$) was tentatively detected with an S/N ratio of 3 in L1521B.

\subsection{Analyses}

\floattable
\begin{deluxetable}{lcc}
\tabletypesize{\scriptsize}
\tablecaption{Column densities and fractional abundances of CCH, column densities of its $^{13}$C isotopologues and $^{12}$C/$^{13}$C ratios in L1521B and L134N \label{tab:t2} }
\tablewidth{0pt}
\tablehead{
\colhead{Parameters} & \colhead{L1521B} & \colhead{L134N}
}
\startdata
$N$(CCH) [cm$^{-2}$]\tablenotemark{a} & ($1.66 \pm 0.18$)$\times 10^{14}$ & ($7.3 \pm 0.9$)$\times 10^{13}$ \\
     $N$($^{13}$CCH) [cm$^{-2}$] & $< 6.2 \times 10^{11}$ & $< 5.3 \times 10^{11}$ \\
     $N$(C$^{13}$CH) [cm$^{-2}$] \tablenotemark{b} & ($7 \pm 2$)$\times 10^{11}$  & ($7.2 \pm 1.4$)$\times 10^{11}$ \\
     $N$(CCH)/$N$($^{13}$CCH)\tablenotemark{c} & $>271$ & $>142$ \\
     $N$(CCH)/$N$(C$^{13}$CH) \tablenotemark{c} & $252^{+77}_{-48}$ & $ 101^{+24}_{-16}$ \\
     $N$(H$_{2}$) [cm$^{-2}$] & $9.5 \times 10^{21}$ & $1.2 \times 10^{22}$ \\
     $X$(CCH) & ($1.75 \pm 0.19$)$\times 10^{-8}$ & ($6.1 \pm 0.8$)$\times 10^{-9}$ \\
 \enddata
\tablecomments{The errors indicate the standard deviation.}
\tablenotetext{a}{The values were derived assuming that $n_{\rm {H_{2}}}$ is $1 \times 10^{5}$ cm$^{-3}$ with the non-LTE method.}
\tablenotetext{b}{The errors were calculated from the errors in integrated intensity. We calculated the errors in integrated intensity from the following equation: $\Delta T_{\rm {mb}} \times \sqrt{n} \times \Delta v$, where $\Delta T_{\rm {mb}}$, $n$, and $\Delta v$ are rms noise level, the number of channel, and the velocity resolution, respectively.}
\tablenotetext{c}{We assumed that the absolute calibration error is cancelled out because all of the lines were observed simultaneously when we derive their errors. This assumption means that the errors of the $^{12}$C/$^{13}$C ratios would not be affected by the absolute calibration error.}
\end{deluxetable}	

We derived the column densities and excitation temperatures of the normal species with the non-LTE code RADEX \citep{2007A&A...468..627V}.
The gas kinetic temperature is assumed to be 10 K, which is a typical value in dark clouds \citep{1998ApJ...503..717H}.
The collision rate coefficients were taken from \citet{2012MNRAS.421.1891S}.
We calculated the parameters using two H$_{2}$ densities ($n_{\rm {H_{2}}}$)  in each source.
The assumed H$_{2}$ densities are $1.0 \times 10^{5}$ cm$^{-3}$ \citep{1998ApJ...503..717H} and $5.0 \times 10^{4}$ cm$^{-3}$ \citep{2004ApJ...617..399H} in L1521B, and $1.0 \times 10^{5}$ cm$^{-3}$ \citep{1998ApJ...503..717H} and $2.1 \times 10^{4}$ cm$^{-3}$ \citep{2000ApJ...542..870D} in L134N.

We derived the column densities and excitation temperatures from the intensities of the two weakest hyperfine components by a least-squares method \citep{2010AA...512A..31S}\footnote{The derived excitation temperatures and optical depths of each hyperfine component are summarized in Table \ref{tab:radex} in Appendix \ref{sec:a2}.}.
The derived column densities and excitation temperatures of CCH are ($1.66 \pm 0.18$)$\times 10^{14}$ cm$^{-2}$ and $6.6 \pm 0.7$ K ($1\sigma$) for $n_{\rm {H_{2}}} = 1 \times 10^{5}$ cm$^{-3}$, and ($2.05 \pm 0.2$)$\times 10^{14}$ cm$^{-2}$ and $5.3 \pm 0.6$ K for $n_{\rm {H_{2}}} = 5 \times 10^{4}$ cm$^{-3}$ in L1521B.
For L134N, the column densities and excitation temperatures of CCH were derived to be ($7.3 \pm 0.9$)$\times 10^{13}$ cm$^{-2}$ and $6.4 \pm 0.8$ K for $n_{\rm {H_{2}}} = 1 \times 10^{5}$ cm$^{-3}$, and ($1.36 \pm 0.17$)$\times 10^{14}$ cm$^{-2}$ and $4.1 \pm 0.5$ K for $n_{\rm {H_{2}}} = 2.1 \times 10^{4}$ cm$^{-3}$, respectively.
Because an excitation temperature of $\sim 6.5$ K is consistent with  typical values of carbon-chain molecules in dark clouds \citep{1992ApJ...392..551S} and the critical density of the $N=1-0$ transition of CCH is $1 \times 10^{5}$ cm$^{-3}$ \citep{2017A&A...605L...5K}, we employ the values obtained with $n_{\rm {H_{2}}} = 1 \times 10^{5}$ cm$^{-3}$ in the following sections.
The excitation temperature of $\sim 6.5$ K is lower than the gas kinetic temperature, but such low excitation temperatures have been derived in prestellar cores \citep[e.g.,][]{2009A&A...505.1199P}.

We derived the column densities of the $^{13}$C isotopologues assuming the LTE condition using the following formulae \citep{2016ApJ...817..147T}:
\begin{equation} \label{tau}
\tau = - {\mathrm {ln}} \left[1- \frac{T_{\rm mb} }{J(T_{\rm {ex}}) - J(T_{\rm {bg}})} \right],  
\end{equation}
where
\begin{equation} \label{tem}
J(T) = \frac{h\nu}{k}\Bigl\{\exp\Bigl(\frac{h\nu}{kT}\Bigr) -1\Bigr\} ^{-1},
\end{equation}  
and
\begin{eqnarray} \label{col}
    N = \tau \frac{3h\Delta v}{8\pi ^3 S}\sqrt{\frac{\pi}{4\mathrm {ln}2}}Q\frac{1}{\mu ^2}\frac{1}{J_{\rm {lower}}+1}\exp\Bigl(\frac{E_{\rm {lower}}}{kT_{\rm {ex}}}\Bigr) 
                    \times \Bigl\{1-\exp\Bigl(-\frac{h\nu }{kT_{\rm {ex}}}\Bigr)\Bigr\} ^{-1}.
\end{eqnarray} 
In equation (\ref{tau}), $T_{\rm mb}$ is the peak intensity (Table \ref{tab:t1}) and $\tau$ is the optical depth.
$T_{\rm {ex}}$ and $T_{\rm {bg}}$ are the excitation temperature and the cosmic microwave background temperature (2.73 K), respectively.
We assumed that the excitation temperatures of the $^{13}$C isotopologues of CCH are equal to those of the normal species.
We then used the excitation temperatures of $6.6 \pm 0.7$ K and $6.4 \pm 0.8$ K in L1521B and L134N, respectively.
{\it J}({\it T}) in equation (\ref{tem}) is the effective temperature equivalent to that in the Rayleigh-Jeans law. 
In equation (\ref{col}), {\it N} denotes the column density, $\Delta v$ the line width (FWHM), $S$ the line strength, $Q$ the rotational partition function, $\mu$ the permanent electric dipole moment, and $E_{\rm {lower}}$ the energy of the lower rotational energy level.
The permanent electric dipole moment is 0.769 Debye for both the $^{13}$C isotopologues \citep{1995CPL...244...45W}.
Taking into account the evaluation of the Gaussian fitting, we derived the column densities of C$^{13}$CH from the line of $J=3/2-1/2, F_{1}= 2- 1, F=5/2-3/2$ in L1521B and the line of $J=3/2-1/2, F_{1}= 2- 1, F=3/2-1/2$ in L134N.
In the case of $^{13}$CCH, we derived the upper limits of column density from the $3\sigma$ upper limits of the peak intensities.
We used the average line widths of the normal species: 0.42 km s$^{-1}$ and 0.35 km s$^{-1}$ in L1521B and L134N, respectively.
We summarize the column densities derived in each source in Table \ref{tab:t2}.

Table \ref{tab:t2} summarizes the H$_{2}$ column density, $N$(H$_{2}$), at the observed positions.
We obtained the $N$(H$_{2}$) value in L1521B from the fits file of the column density map\footnote{Taken from \url{http://www.herschel.fr/cea/gouldbelt/en/Phocea/Vie_des_labos/Ast/ast_visu.php?id_ast=66}}.
The map was made using the Herschel data \citep[70, 160, 250, 350, and 500 $\mu$m;][]{2013A&A...550A..38P}.
We derived the H$_{2}$ column density in L134N from the archival data of the 1.2 mm dust continuum emission obtained by the MAMBO bolometer array installed on the IRAM 30 m telescope\footnote{Taken from \url{http://cdsweb.u-strasbg.fr/cgi-bin/qcat?J/A+A/487/993}}, using the following formula \citep{2008A&A...487..993K}:
\begin{equation} \label{H2}
N({\rm{H}}_{2}) = 6.69 \times 10^{20} \times F_{\nu},
\end{equation}
where $F_{\nu}$ is the flux intensity in unit of mJy beam$^{-1}$.
The flux intensity is 17.9 mJy beam$^{-1}$ at the observed position, and the derived H$_{2}$ column density is $1.2 \times 10^{22}$ cm$^{-2}$ using equation (\ref{H2}).
The fractional abundances of CCH, $X$(CCH)$=N$(CCH)/$N$(H$_{2}$), are calculated at ($1.75 \pm 0.19$)$\times 10^{-8}$ and ($6.1 \pm 0.8$)$\times 10^{-9}$ in L1521B and L134N, respectively.

\section{Discussion} \label{sed:dis}

\subsection{$^{13}$C Isotopic Fractionation of CCH in L1521B and L134N} \label{sec:dis1}

\begin{figure}[!th]
\figurenum{4}
 \begin{center}
  \includegraphics[width=11cm, bb= 0 150 612 792]{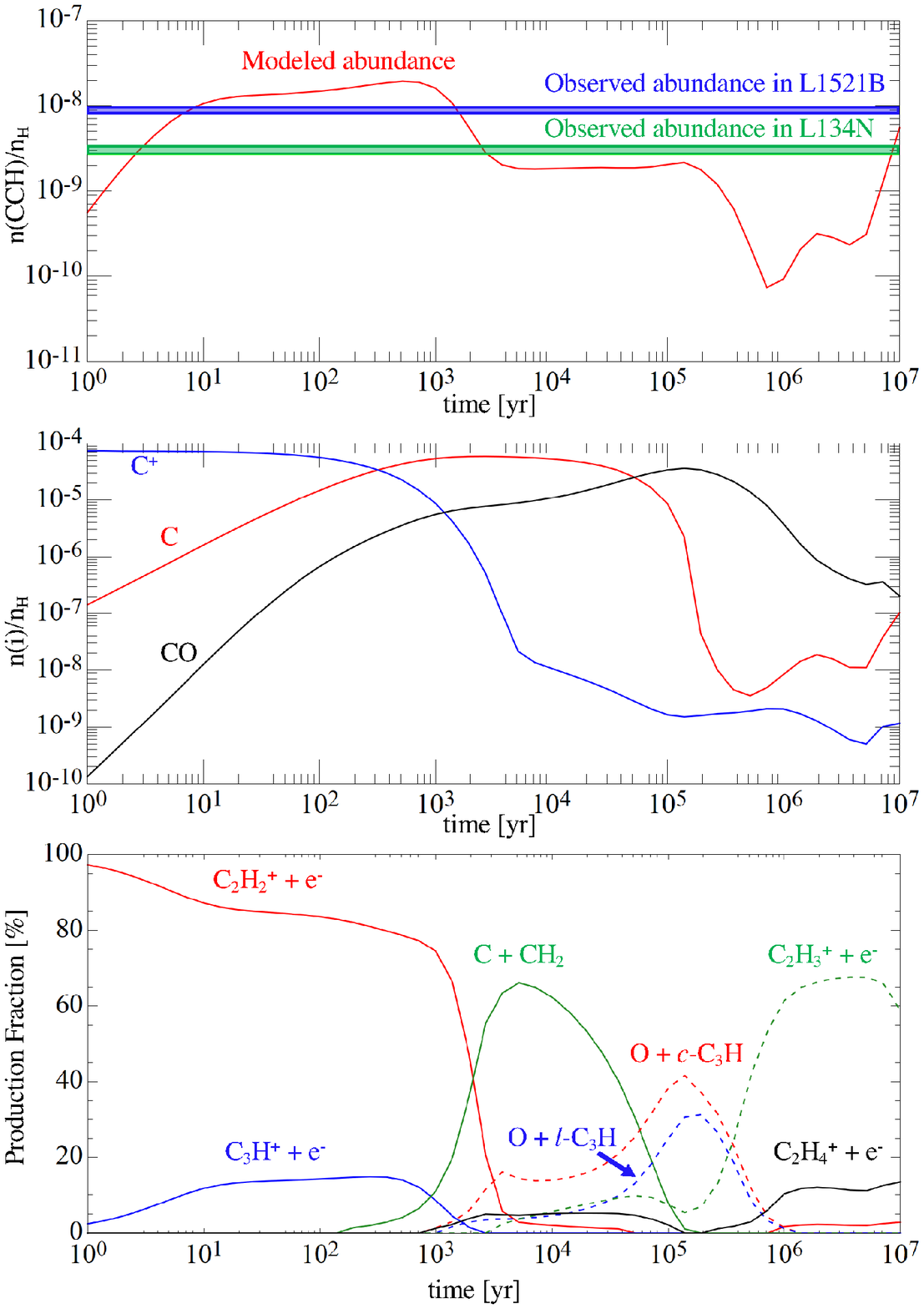} 
 \end{center}
\caption{Results of the model calculation. Upper panel; the time dependence of the CCH abundance respect to total hydrogen (red line). The blue and green lines indicate the observed abundances in L1521B and L134N, respectively. Middle panel; the time dependence of the C (red), C$^{+}$ (blue), and CO (black) abundances respect to hydrogen. Lower panel; the time dependence of the contribution of each reaction to form CCH. \label{fig:f4}}
\end{figure}

In this section, we compare the column densities between the two $^{13}$C isotopologues of CCH, namely the fractionation between the two  $^{13}$C isotopomers, in the observed two starless cores.
Because we could not detect $^{13}$CCH with an S/N ratio above 3 in the two observed starless cores, we derived  lower limits of the $N$(C$^{13}$CH)/$N$($^{13}$CCH) ratio, which are $>1.1$ and $>1.4$ in L1521B and L134N, respectively\footnote{These lower limits have the errors of 0.3, which are derived from the standard deviation of $N$(C$^{13}$CH).}.
In both sources, the $^{13}$CCH isotopomer is less abundant than the  C$^{13}$CH isotopomer.
This result  is the same as found for TMC-1 and L1527 by \citet{2010AA...512A..31S}, where the  C$^{13}$CH/$^{13}$CCH abundance ratios were  derived to be $1.6 \pm 0.4$ and $1.6 \pm 0.1$ ($3 \sigma$) respectively.
Based on these results, it may be common that $^{13}$CCH tends to be less abundant in starless cores.

Two possible mechanisms causing the $^{13}$C isotopomer fractionation in CCH have been  proposed: fractionation via the formation pathway \citep{2010AA...512A..31S} and via an isotopomer-exchange reaction \citep{2011ApJ...731...38F}.
\citet{2010AA...512A..31S} discussed the formation pathways of CCH that could cause its $^{13}$C isotopic fractionation.
They considered the following three reactions:
\begin{equation} \label{rea:r1}
{\rm {C}}_{2}{\rm {H}}_{2}^{+} + {\rm {e}}^{-} \rightarrow {\rm {C}}_{2}{\rm {H}} + {\rm {H}},
\end{equation}
\begin{equation} \label{rea:r2}
{\rm {C}}_{2}{\rm {H}}_{3}^{+} + {\rm {e}}^{-} \rightarrow {\rm {C}}_{2}{\rm {H}} + {\rm {H}}_{2},
\end{equation}
and 
\begin{equation} \label{rea:r3}
{\rm {C}}{\rm {H}}_{2} + {\rm {C}} \rightarrow {\rm {C}}_{2}{\rm {H}} + {\rm {H}}.
\end{equation}
Among the above three reactions, only reaction (\ref{rea:r3}) is able to cause the $^{13}$C isotopic fractionation in CCH, because the two carbon atoms are not clearly equivalent.
Hence, \citet{2010AA...512A..31S} deduced that the observed differences in the abundances between $^{13}$CCH and C$^{13}$CH would reflect the significant contribution of reaction (\ref{rea:r3}).
However, the contributions of each reaction to the overall formation of CCH were not investigated in detail.

We chose to calculate the contributions of each formation pathway of CCH using the astrochemical code Nautilus \citep{2016MNRAS.459.3756R}.  
Our model calculation and the reaction network utilized  are described in Appendix \ref{sec:a1}.
Figure \ref{fig:f4} shows the results of the model calculation.
The upper panel shows  the CCH abundance with respect to total hydrogen as a function of time, as well as horizontal lines for the observed abundances in L1521B and L134N.   
Given the standard level of agreement between calculated and observed abundances in dark clouds,  it can be argued that the modeled abundance shows substantial agreement with the observed abundances in both sources over significant periods of time.  In addition, estimates of dark cloud ages based on the agreement between observed and calculated abundances for large numbers of molecular species indicate a much tighter constraint on ages centered on the so-called ``early time'' of $\approx  10^{5}$ yr \citep{2006A&A...451..551W}. 

The lower panel of Figure \ref{fig:f4} shows the contribution of each major reaction to the rate of formation of CCH.
We exclude reactions which have fractions below 10\%.  
Before $10^{3}$ yr, reaction (\ref{rea:r1}) is the major formation pathway of CCH.
The C$_{2}$H$_{2}^{+}$ ion has two equivalent carbon atoms, so this reaction cannot explain the differences in abundances between the two $^{13}$C isotopologues of CCH.
The following reaction has the second highest contribution in this time range: 
\begin{equation} \label{rea:r4}
{\rm {C}}_{3}{\rm {H}}^{+} + {\rm {e}}^{-} \rightarrow {\rm {C}}_{2}{\rm {H}} + {\rm {C}}.
\end{equation}
The three carbon atoms in C$_{3}$H$^{+}$ are not equivalent, and this can explain the $^{13}$C isotopic fractionation, unless scrambling of the carbon atoms occurs efficiently during the electron recombination reaction.
However, this reaction contributes a  significantly smaller amount to the formation of CCH  than reaction (\ref{rea:r1}),  and we cannot conclude that reaction (\ref{rea:r4}) significantly contributes to the observable $^{13}$C isotopic fractionation of CCH.
In addition, although unlikely in such a small system, a scrambling of carbon atoms may occur during ion-molecule reaction that produces C$_{3}$H$^{+}$ \citep{2016ApJ...817..147T}.
In that case, we would not recognize clear differences in abundances among the $^{13}$C isotopologues.
The contribution of reaction (\ref{rea:r4}) to the $^{13}$C isotopic fractionation of CCH is still unclear.

After $10^{3}$ yr, reaction (\ref{rea:r3}) has the largest contribution to the formation of CCH. 
As first explained by \citet{2010AA...512A..31S}, this reaction can result in $^{13}$C fractionation because the carbon atoms are not identical.  
Another way of looking at the problem is that unless the insertion of a carbon atom into the C-H bond occurs at the same rate as its addition to the C of CH$_{2}$ and the C atoms can scramble, fractionation will occur.   
The dominant contribution of reaction (\ref{rea:r3}) extends from somewhat greater than 10$^{3}$ yr to almost 10$^{5}$ yr, which is a much longer period than the time range when reaction (\ref{rea:r1}) is dominant. 
Moreover, its range of dominance coincides more closely with the age range determined in multi-molecule fits to abundances in dark clouds \citep{2006A&A...451..551W}. 
This reaction should therefore contribute to  the differences in abundances between $^{13}$CCH and C$^{13}$CH, as mentioned before.

Now consider the case of the  isotopomer-exchange reaction:
\begin{equation}
 {\rm ^{13}C^{12}CH + H \rightleftharpoons ^{12}C^{13}CH + H }+ \Delta E (8.1 K), 
 \end{equation}
 which is exothermic as written from left-to-right so that the net effect is to increase the abundance of the  $^{12}$C$^{13}$CH isotopomer at the expense of $^{13}$C$^{12}$CH isotopomer.   
\citet{2011ApJ...731...38F} used $1\times 10^{-10}$ cm$^{3}$ s$^{-1}$ as the rate coefficient of the forward reaction of the isotopomer-exchange reaction in their calculation.
They also applied 8.1 K to the zero-point energy difference of $^{13}$CCH and C$^{13}$CH.
The efficiency of this process depends upon a number of factors including whether the exothermicity of 8.1 K is sufficient to cause the difference in observed abundance, whether the abundance of atomic hydrogen is large enough, and whether or not a barrier to reaction exists.  The  differences in abundances between $^{13}$CCH and C$^{13}$CH can be seen even at the early stages \citep[Figure 3 in][]{2011ApJ...731...38F}.
This isotopomer-exchange reaction appears to be able to explain at least partially the $^{13}$C isotopic fractionation in both L1521B and L134N.

In summary, we found that C$^{13}$CH is more abundant than $^{13}$CCH in L1521B and L134N from our observations.
This tendency agrees with the previous observations in TMC-1 and L1527 \citep{2010AA...512A..31S}.
The higher abundance of C$^{13}$CH compared with $^{13}$CCH may indeed be common for dark clouds.
Both reaction (\ref{rea:r3}) and the isotopomer-exchange reaction likely contribute to the $^{13}$C isotopic fractionation of CCH both in L1521B and L134N.

\subsection{The Dilution of the $^{13}$C Species in Dark Clouds} \label{sec:dis2}

\subsubsection{Comparisons of the $^{12}$C/$^{13}$C Ratios between CCH and HC$_{3}$N among Dark Clouds} \label{sec:dis2_1}

\floattable
\begin{deluxetable}{lccc}
\tabletypesize{\scriptsize}
\tablecaption{The $^{12}$C/$^{13}$C ratios of CCH and HC$_{3}$N in dark clouds \label{tab:frac}}
\tablewidth{0pt}
\tablehead{
\colhead{Species} &  \colhead{L1521B} & \colhead{L134N} & \colhead{TMC-1}
}
\startdata
$^{13}$CCH & $>271$ & $>142$ & $>250$\tablenotemark{a} \\
C$^{13}$CH & $252^{+77}_{-48}$ & $101^{+24}_{-16}$ & $>170$\tablenotemark{a} \\
H$^{13}$CCCN & $117 \pm 16$\tablenotemark{b} & $61 \pm 9$\tablenotemark{b} & $79 \pm 11$\tablenotemark{c} \\
HC$^{13}$CCN & $115 \pm 16$\tablenotemark{b} & $94 \pm 26$\tablenotemark{b} & $75 \pm 10$\tablenotemark{c} \\
HCC$^{13}$CN & $76 \pm 6$\tablenotemark{b} & $46 \pm 9$\tablenotemark{b} & $55 \pm 7$\tablenotemark{c} \\
\enddata
\tablenotetext{a}{Taken from \citet{2010AA...512A..31S}.}
\tablenotetext{b}{Taken from \citet{2017ApJ...846...46T}.}
\tablenotetext{c}{Taken from \citet{1998AA...329.1156T}.}
\end{deluxetable}	

\begin{figure}[!th]
\figurenum{5}
 \begin{center}
  \includegraphics[width=13cm, bb= 0 0 428 214]{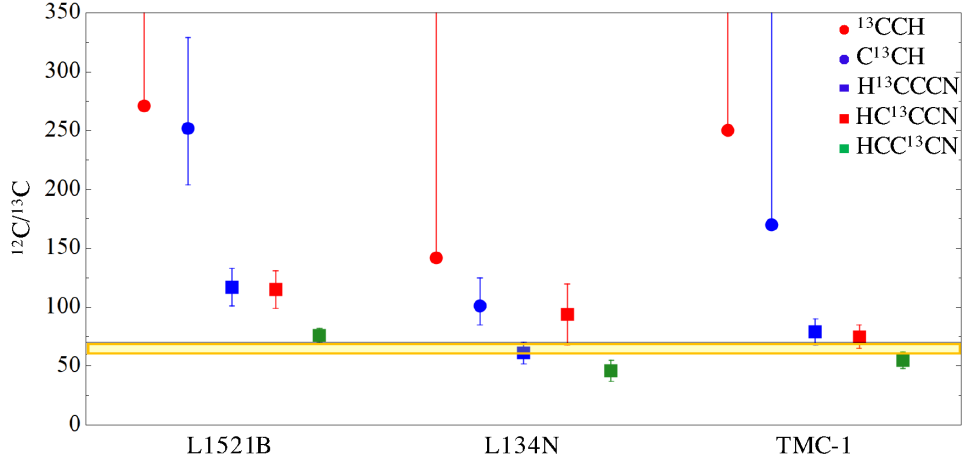} 
 \end{center}
\caption{Comparisons of the $^{12}$C/$^{13}$C ratios of CCH and HC$_{3}$N in the three starless cores; L1521B, L134N, and TMC-1. The errors indicate the standard deviation. The yellow range ($^{12}$C/$^{13}$C$= 60-70$) shows the values in the local interstellar medium. The values are the same as Table \ref{tab:frac}. \label{fig:f5}}
\end{figure}

It is interesting to compare the $^{12}$C/$^{13}$C ratios of CCH and HC$_{3}$N among the three dark clouds L1521B, L134N, and TMC-1.
Table \ref{tab:frac} summarizes the $^{12}$C/$^{13}$C ratios of CCH and HC$_{3}$N in the three dark clouds and Figure \ref{fig:f5} shows the comparisons.
In the local interstellar medium, the elemental $^{12}$C/$^{13}$C ratio has been determined to be $60-70$ \citep[e.g.,][]{2005ApJ...634.1126M}, which we indicate as the yellow range in Figure \ref{fig:f5}.

From Figure \ref{fig:f5}, we find that the dilution of the $^{13}$C species in carbon-chain molecules holds for the three observed dark clouds.
In addition, the $^{12}$C/$^{13}$C ratios of CCH tend to be higher than those of HC$_{3}$N in all of the dark clouds.
It had already been suggested that the degrees of the dilution are different among carbon-chain species in TMC-1 \citep{2016ApJ...817..147T}.
We now can confirm the different degrees of the dilution among the carbon-chain species in the other dark clouds from our observations.

As mentioned in Section \ref{sec:intro}, the reaction between CCH and HNC was proposed as the main formation pathway of HC$_{3}$N in L134N \citep{2017ApJ...846...46T}.
In this reaction, the carbon atom in HNC attacks the carbon atom with an unpaired electron in CCH forming HC$_{3}$N via HCCCNH \citep{1997ApJ...489..113F}.
In that case, we can distinguish all the carbon atoms and trace them during the reaction scheme.
We would expect that HC$_{3}$N/HC$^{13}$CCN is equal to CCH/$^{13}$CCH and HC$_{3}$N/H$^{13}$CCCN is equal to CCH/C$^{13}$CH. 
We indicated for the former pair (HC$^{13}$CCN/HC$_{3}$N and $^{13}$CCH/CCH) with red symbols and for the latter pair with blue symbols in Figure \ref{fig:f5}.
The latter pair agree within their $2\sigma$ error bars (CCH/C$^{13}$CH $=69-149$ and HC$_{3}$N/H$^{13}$CCCN $=43-79$) taking the $2\sigma$ errors into consideration.
The former pair may marginally lie within the $2\sigma$ error bars (CCH/$^{13}$CCH $>142$ and HC$_{3}$N/HC$^{13}$CCN $=42-146$), taking the $2\sigma$ error into consideration, but we cannot strongly confirm it due to the non-detection of $^{13}$CCH.
We need data with higher sensitivity in order to reach this conclusion.

In L1521B and TMC-1, the $^{12}$C/$^{13}$C ratios of CCH are higher than those of HC$_{3}$N by more than a factor of 2.
The reaction between C$_{2}$H$_{2}$ and CN was proposed as the main formation pathway of HC$_{3}$N in both the dark clouds \citep{1998AA...329.1156T,2017ApJ...846...46T}, and the carbon atom which is next to the nitrogen atom in HC$_{3}$N should originate from CN, while the other two carbon atoms in HC$_{3}$N should come from C$_{2}$H$_{2}$ \citep{1997ApJ...489..113F}.
Taking these points into consideration, the $^{12}$C/$^{13}$C ratios of H$^{13}$CCCN and HC$^{13}$CCN could reflect those of C$_{2}$H$_{2}$.
The C$_{2}$H$_{2}$ molecule is mainly formed by the electron recombination reaction of C$_{2}$H$_{4}^{+}$, which is formed by the reaction between C$_{2}$H$_{2}^{+}$ and H$_{2}$ (Figure \ref{fig:f6}).
The C$_{2}$H$_{2}^{+}$ ion also forms CCH via its electron recombination reaction. 
Hence, the $^{12}$C/$^{13}$C ratio of C$_{2}$H$_{2}$ is expected to be similar to those of CCH, if the $^{12}$C/$^{13}$C ratios are determined during their bottom-up formation from C$^{+}$ and/or C.
However, the observational results show discrepancies in the $^{12}$C/$^{13}$C ratio between CCH and C$_{2}$H$_{2}$.
This suggests that the different degrees of the dilution of the $^{13}$C species are not induced during the carbon-chain growth from C$^{+}$ and/or C, but during reactions occurring after their production.
In the following subsection, we discuss possible routes which could cause the significantly high $^{12}$C/$^{13}$C ratios of CCH, especially at a young dark cloud stage.

\subsubsection{Possible Routes to Produce the Significantly High $^{12}$C/$^{13}$C Ratios in CCH} \label{sec:dis2_2}

\begin{figure}[!th]
\figurenum{6}
 \begin{center}
  \includegraphics[width=14cm, bb= 0 0 791 425]{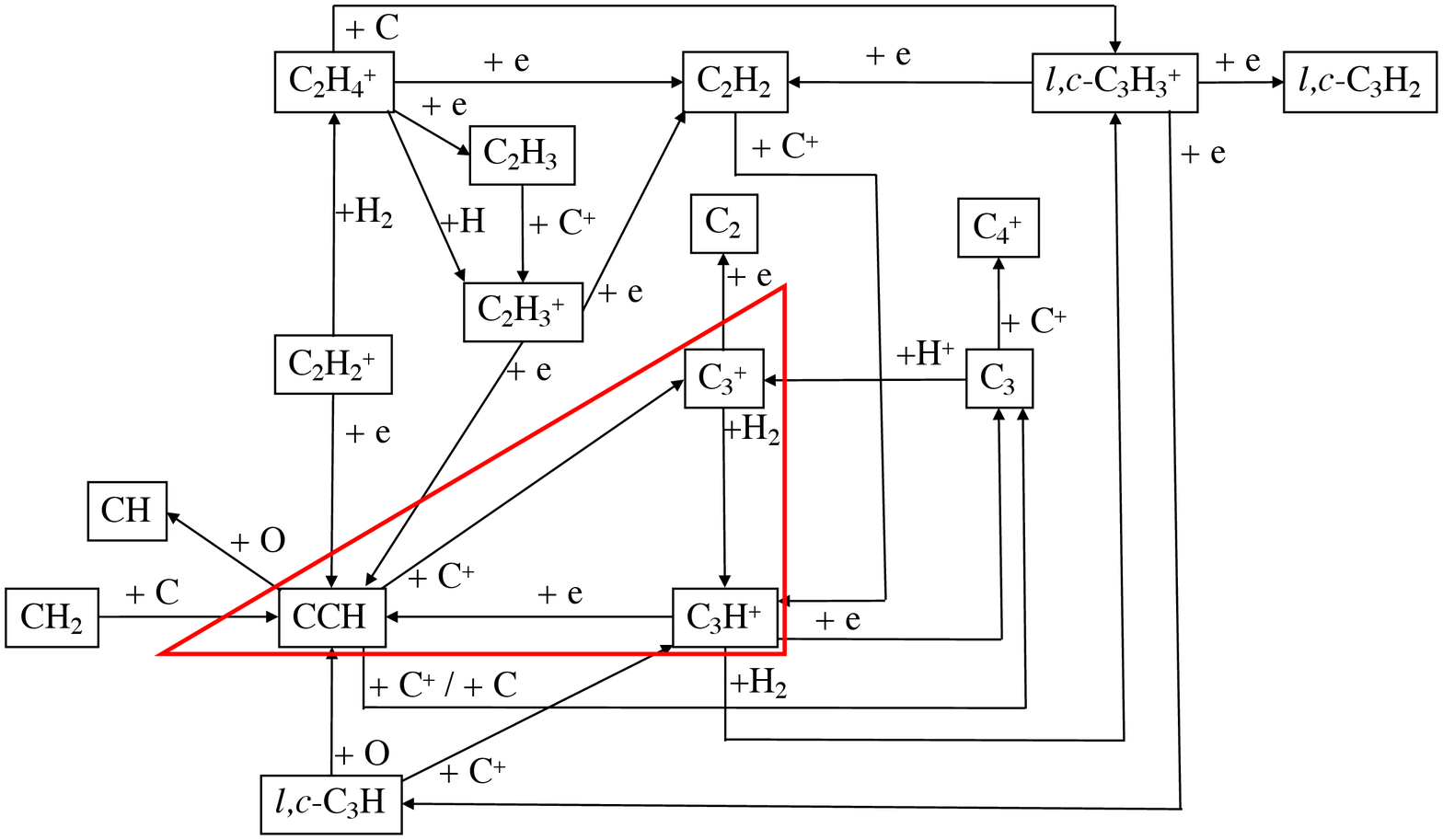} 
 \end{center}
\caption{Reaction schemes of small hydrocarbons. A red triangle highlights the reaction cycle that causes the high $^{12}$C/$^{13}$C ratios in CCH and is efficient especially in the early stage ($t < 10^{3}$ yr). \label{fig:f6}}
\end{figure}

From our model calculation, we uncovered efficient formation and destruction pathways of small hydrocarbons as shown in Figure \ref{fig:f6}.
These pathways were found as we searched for possible routes which cause the high $^{12}$C/$^{13}$C ratio  in CCH.  

An important cycle, which seems to be efficient in increasing the $^{12}$C/$^{13}$C ratio in CCH in stages as early as $t < 10^{3}$ yr, is  highlighted as a red triangle in Figure \ref{fig:f6}.
In this cycle, more than 80\% of CCH is destroyed by the reaction with C$^{+}$ to form C$_{3}^{+}$ before $10^{3}$ yr. 
CCH is reformed by the reaction of C$_{3}^{+}$ with H$_{2}$ followed by the dissociative recombination of C$_{3}$H$^{+}$.
If $^{13}$C$^{+}$ is diluted due to reaction (\ref{rea:co}), the $^{12}$C/$^{13}$C ratios of CCH will become higher during the cycle because this cycle involves $^{13}$C$^{+}$.
Another possible route to produce the high $^{12}$C/$^{13}$C ratio  in CCH involves the following reaction:
\begin{equation}
{\rm {C}}_{2}{\rm {H}}_{2} + {\rm {C}}^{+} \rightarrow {\rm {C}}_{3}{\rm {H}}^{+} + {\rm {H}}.
\end{equation}
The C$_{3}$H$^{+}$ product once again reacts dissociatively with electrons to form CCH + C.
If the $^{12}$C/$^{13}$C ratio of C$_{3}$H$^{+}$ is high, the ratio of CCH also should be high.
The following ion-neutral bimolecular reaction could transfer the high $^{12}$C/$^{13}$C ratio to C$_{3}$H$^{+}$: 
\begin{equation}
c,l-{\rm {C}}_{3}{\rm {H}} + {\rm {C}}^{+} \rightarrow {\rm {C}}_{3}{\rm {H}}^{+} + {\rm {C}}.
\end{equation}
The high $^{12}$C/$^{13}$C ratio of C$_{3}$H$^{+}$ may increase the $^{12}$C/$^{13}$C ratio of $l,c$-C$_{3}$H$_{3}^{+}$. 
However,  most electron recombination reactions of $l,c$-C$_{3}$H$_{3}^{+}$ lead to the formation of $l,c$-C$_{3}$H$_{2}$ and $l,c$-C$_{3}$H.
In that case, the high $^{12}$C/$^{13}$C ratio of $l,c$-C$_{3}$H$_{3}^{+}$ will not significantly affect the ratio of C$_{2}$H$_{2}$.
Therefore, the high $^{12}$C/$^{13}$C ratio of C$_{3}$H$^{+}$ will not significantly affect the ratio of C$_{2}$H$_{2}$, but produce the high ratio of CCH.

\citet{2011ApJ...731...38F} plotted the temporal variation of the $^{12}$C$^{+}$/$^{13}$C$^{+}$ ratio.
In fact, this ratio takes extremely high value of $\geq450$ around $t  \simeq 10^{3}$ yr with the assumed density of $5 \times 10^{4}$ cm$^{-3}$ \citep[Figure 1 in][]{2011ApJ...731...38F}.
Therefore, there is a possibility that the above reactions cause the significantly high $^{12}$C/$^{13}$C ratios in CCH at an early time.

After $t = 10^{3}$ yr, the abundance of ionic carbon (C$^{+}$) rapidly decreases (see the middle panel of Figure \ref{fig:f4}), and the reactions with C$^{+}$ are suppressed. 
In addition, the $^{12}$C$^{+}$/$^{13}$C$^{+}$ ratio becomes lower in the later stage \citep[$^{12}$C$^{+}$/$^{13}$C$^{+}$$\approx 200$ at $2 \times 10^{3} < t < 3 \times 10^{4}$ yr;][]{2011ApJ...731...38F}.
Hence, the $^{12}$C/$^{13}$C ratios of CCH would not increase due to the above reactions.
These model results may support the lower $^{12}$C/$^{13}$C ratios of CCH in L134N compared to younger clouds of L1521B and TMC-1\citep{2004ApJ...617..399H,2000ApJ...542..870D,2017ApJ...846...46T}.

\citet{2011ApJ...731...38F} computed and displayed the temporal variation of the $^{12}$C/$^{13}$C ratio of CCH.
The modeled ratio takes its peak value of $\sim220$ at a  time around $3 \times 10^{2}$ yr and quickly decreases to 100 -- 60.
The predicted peak value is still lower than the observed value in L1521B, and their prediction is not consistent with the observational results quantitatively.
However, the observed results that the $^{12}$C/$^{13}$C ratios of CCH in the chemically evolved dark cloud (L134N) are lower compared to the chemically young dark clouds, L1521B and TMC-1, are qualitatively consistent with the simulation \citep{2011ApJ...731...38F}.
As they pointed out, \citet{2011ApJ...731...38F} may lack some mechanisms which cause the dilution of the $^{13}$C species.
For example, \citet{2011ApJ...731...38F} did not take the selective photodissociation into consideration, but it will increase the $^{12}$C/$^{13}$C ratios.
If the selective photodissociation of CCH occurs, the $^{13}$C isotopologues are destroyed in denser regions where the normal species can survive.
CCH seems to be optically thick because the optical thickness of the weakest hyperfine components are around 0.2 in L1521B.

\section{Conclusions} \label{sec:con}

We have carried out observations of the $N=1-0$ transition lines of CCH and its two $^{13}$C isotopologues toward two starless cores, L1521B and L134N, using the Nobeyama 45 m radio telescope.
The isotopologue of C$^{13}$CH is detected with an S/N ratio of 4, while the other isotopologue, $^{13}$CCH, could not be detected with an S/N ratio above 3.
The $N$(C$^{13}$CH)/$N$($^{13}$CCH) ratios are derived to be $>1.1$ and $>1.4$ in L1521B and L134N, respectively.
The characteristic that C$^{13}$CH is more abundant than $^{13}$CCH seems to be common for cold dark clouds.
Such a difference in abundances between the two $^{13}$C isotopologues, namely the $^{13}$C isotopic fractionation, of CCH could be caused during its formation pathway and by the isotopomer-exchange reaction after the molecule is formed.

The derived $^{12}$C/$^{13}$C ratios of CCH in L1521B and L134N are higher than the elemental ratio in the local interstellar medium.
We compared the $^{12}$C/$^{13}$C ratios of CCH and HC$_{3}$N among the three dark clouds.
The $^{12}$C/$^{13}$C ratios of CCH are higher than those of HC$_{3}$N by more than a factor of 2 in L1521B and TMC-1, while the differences in the $^{12}$C/$^{13}$C ratios between CCH and HC$_{3}$N seem to be smaller in L134N.
We discussed possible routes to produce the significantly high $^{12}$C/$^{13}$C ratios only in CCH based on the chemical network simulation.
We found a possible cycle which occurs efficiently in the early stage of dark clouds.
The previous study shows that the  $^{12}$C$^{+}$/$^{13}$C$^{+}$ ratio becomes extremely high above 450 in the early time of dark cloud \citep{2011ApJ...731...38F}.
Taking the predicted $^{12}$C$^{+}$/$^{13}$C$^{+}$ into account, the $^{12}$C/$^{13}$C ratio of CCH will also become high in the cycle because this cycle involves C$^{+}$.
Besides, the reactions of ``C$_{2}$H$_{2}$ + C$^{+}$" and ``$l,c$-C$_{3}$H + C$^{+}$" can contribute to the high $^{12}$C/$^{13}$C ratios in CCH.

\acknowledgments
We are deeply grateful to the staff of the Nobeyama Radio Observatory.
The Nobeyama Radio Observatory is a branch of the National Astronomical Observatory of Japan (NAOJ), National Institutes of Natural Science (NINS).
K. T. would like to thank the University of Virginia for providing the funds for her postdoctoral fellowship in the Virginia Initiative on Cosmic Origins (VICO) research program.
E. H. would like to thank the National Science Foundation for support of his program in astrochemistry.

%

\vspace{5mm}
\facilities{Nobeyama 45 m radio telescope}


\software{RADEX \citep{2007A&A...468..627V}, Nautilus \citep{2016MNRAS.459.3756R}}



\appendix

\section{Model Calculation} \label{sec:a1}

We calculated the abundance of CCH and the contributions of each formation/destruction pathway using the astrochemical code Nautilus \citep{2016MNRAS.459.3756R}.
The initial elemental abundances with respect to total hydrogen are taken from \citep{2017ApJ...850..105A} as summarized in Table \ref{tab:ie}.
Initially, all of hydrogen is the form in H$_{2}$.
The initial form of hydrogen does not affect our discussion.
We ran the model calculation including 7646 gas-phase reactions and 498 gas-phase species,  mainly taken from the Kinetic Database for Astrochemistry (KIDA)\footnote{http://kida.obs.u-bordeaux1.fr}.
There are 5323 grain-surface reactions and 431 grain-surface species including suprathermal species \citep{2018PCCP...20.5359S}.
The surface reactions come mainly from \citet{2013ApJ...765...60G}, with additional data taken from \citet{2018ApJ...852...70B} and \citet{2018ApJ...857...89H}.
The self shielding effects of H$_{2}$ \citep{1996A&A...311..690L}, CO \citep{2009A&A...503..323V}, and N$_{2}$ \citep{2013A&A...555A..14L} are included.  

The assumed density, gas temperature, visual extinction ($A_{\rm {v}}$), and cosmic-ray ionization rate ($\zeta$) are $2 \times 10^{4}$ cm$^{-3}$, 10 K, 10 mag, and $1.3 \times 10^{-17}$ s$^{-1}$, respectively.
We assume that the dust temperature is equal to the gas temperature.
These values are considered to be the typical values for dark clouds \citep{2013A&A...550A..36M}.

\floattable
\begin{deluxetable}{cc}
\tabletypesize{\scriptsize}
\tablecaption{Initial elemental abundances with respect to total hydrogen \label{tab:ie}}
\tablewidth{0pt}
\tablehead{
\colhead{Element} &  \colhead{Abundance}
}
\startdata
H$_{2}$ & 0.5 \\
He & 0.09 \\
C$^{+}$ & $7.3 \times 10^{-5}$ \\
N & $2.14 \times 10^{-5}$ \\
O & $1.76 \times 10^{-4}$ \\
F & $1.8 \times 10^{-8}$ \\
Si$^{+}$ & $8 \times 10^{-9}$ \\
S$^{+}$ & $8 \times 10^{-8}$ \\
Fe$^{+}$ & $3 \times 10^{-9}$ \\
Na$^{+}$ & $2 \times 10^{-9}$ \\
Mg$^{+}$ & $7 \times 10^{-9}$ \\
Cl$^{+}$ & $1 \times 10^{-7}$ \\
P$^{+}$ & $2 \times 10^{-10}$ \\
\enddata
\tablecomments{Taken from the AL model in \citet{2017ApJ...850..105A}.}
\end{deluxetable}	

\section{Excitation temperature and optical depth of each hyperfine component} \label{sec:a2}

Table \ref{tab:radex} summarizes the excitation temperatures and optical depths of CCH of each hyperfine component derived by the RADEX.

\floattable
\begin{deluxetable}{lcccccccccc}
\tabletypesize{\scriptsize}
\tablecaption{Excitation temperature and optical depth of CCH of each hyperfine component \label{tab:radex}}
\tablewidth{0pt}
\tablehead{
\colhead{} &  \colhead{} & \multicolumn{4}{c}{L1521B} & \colhead{} & \multicolumn{4}{c}{L134N} \\
\cline{3-6} \cline{8-11} 
\colhead{Line} & \colhead{$S$} & \colhead{$T_{\rm {ex}}$ (K)\tablenotemark{a} }& \colhead{$\tau$\tablenotemark{a}} & \colhead{$T_{\rm {ex}}$ (K)\tablenotemark{b}}& \colhead{$\tau$\tablenotemark{b}} & & \colhead{$T_{\rm {ex}}$ (K)\tablenotemark{a}}& \colhead{$\tau$\tablenotemark{a}} & \colhead{$T_{\rm {ex}}$(K)\tablenotemark{c}}& \colhead{$\tau$\tablenotemark{c}}
}
\startdata
$J=3/2-1/2, F= 1- 1$ & 0.17 & 6.8 & 0.16 & 5.4 & 0.26 & & 6.6 & 0.10 & 4.1 & 0.33 \\
$J=3/2-1/2, F= 2- 1$ & 1.67 & 6.5 & 0.38 & 5.1 & 0.73 & & 6.4 & 0.27 & 4.0 & 1.30 \\
$J=3/2-1/2, F= 1- 0$ & 0.83 & 6.3 & 0.27 & 4.9 & 0.52 & & 6.3 & 0.20 & 3.8 & 0.95 \\
$J=1/2-1/2, F= 1- 1$ & 0.83 & 6.2 & 0.40 & 5.0 & 0.75 & & 6.1 &0.27 & 3.9 & 1.21 \\
$J=1/2-1/2, F= 0- 1$ & 0.33 & 6.3 & 0.29 & 5.0 & 0.51 & & 6.2 & 0.17 & 3.9 & 0.70 \\
$J=1/2-1/2, F= 1- 0$ & 0.17 & 6.5 & 0.20 & 5.3 & 0.33 & & 6.3 & 0.13 & 4.1 & 0.41 \\
\enddata
\tablenotetext{a}{$n_{\rm {H_{2}}}=1.0 \times 10^{5}$ cm$^{-3}$.}
\tablenotetext{b}{$n_{\rm {H_{2}}}=5.0 \times 10^{4}$ cm$^{-3}$.}
\tablenotetext{c}{$n_{\rm {H_{2}}}=2.1 \times 10^{4}$ cm$^{-3}$.}
\end{deluxetable}	

\end{document}